\documentclass[12pt]{iopart}

\usepackage[breaklinks=true,colorlinks=true,backref,pagebackref]{hyperref} 

\usepackage{amssymb}
\usepackage{color}

\def\a{\alpha}
\def\b{\beta}
\def\g{\gamma}
\def\d{\delta}

\def\vt{\vartheta}
\def\r{\rho}
\def\l{\lambda}

\def\semidirect{\;{\rlap{$\supset$}\times}\;}

\def\rstar {{\color{red}(*)}}

\begin{document}
\title[Beyond Einstein-Cartan gravity...]{Beyond Einstein-Cartan
  gravity: Quadratic torsion and curvature invariants with even and
  odd parity including all boundary terms}

\author{Peter Baekler$^1$ and Friedrich W Hehl$^{2,3}$}

\address{$^1$ Department of Media, Fachhochschule D\"usseldorf,
  University of Applied Sciences, 40474 D\"usseldorf, Germany}
\address{$^2$ Institute for Theoretical Physics, University of
  Cologne, 50923 K\"oln, Germany} \address{$^3$ Department of Physics
  and Astronomy, University of Missouri, Columbia, MO 65211, USA}
\ead{peter.baekler@fh-duesseldorf.de, hehl@thp.uni-koeln.de (corr.\
  author) {\it file {BeyondEC22.tex, 14 October 2011}}}

\begin{abstract}
  Recently, gravitational gauge theories with torsion have been
  discussed by an increasing number of authors {}from a classical as
  well as {}from a quantum field theoretical point of view. The
  Einstein-Cartan(-Sciama-Kibble) Lagrangian has been enriched by the
  parity odd pseudoscalar curvature (Hojman, Mukku, and Sayed) and by
  torsion square and curvature square pieces, likewise of even and odd
  parity. (i) We show that the inverse of the so-called
  Barbero-Immirzi parameter multiplying the pseudoscalar curvature,
  because of the topological Nieh-Yan form, can be appropriately
  discussed if torsion square pieces are included. (ii) The quadratic
  gauge Lagrangian with both parities, proposed by Obukhov et al.\ and
  Baekler et al., emerges also in the framework of Diakonov et al.\
  (2011). We establish the exact relations between both approaches by
  applying the topological Euler and Pontryagin forms in a
  Riemann-Cartan space expressed for the first time in terms of
  irreducible pieces of the curvature tensor. (iii) In a
  Riemann-Cartan spacetime, that is, in a spacetime with torsion,
  parity violating terms can be brought into the gravitational
  Lagrangian in a straightforward and natural way. Accordingly,
  Riemann-Cartan spacetime is a natural habitat for chiral fermionic
  matter fields.
\end{abstract}

%\submitto{Fast Track Communication \CQG}
\maketitle

%%%%%%%%%%%%%%%%%%%%%%%%%%%%%%%%%%%%%%%%%%%%%%%%%%%%%%%%%%%%%%%%%%%%%
\section{Einstein-Cartan theory and weak gravity}
%%%%%%%%%%%%%%%%%%%%%%%%%%%%%%%%%%%%%%%%%%%%%%%%%%%%%%%%%%%%%%%%%%%%%

In gauge-theoretical approaches to gravity (see
\cite{Erice95,Blagojevic,Ortin,Obukhov:2006gea}), we have the
orthonormal coframe 1-form $\vt^\a$ as the translational potential and
the connection 1-form $\Gamma^{\a\b}=-\Gamma^{\b\a}$ as the Lorentz
potential. The corresponding field strengths are the torsion 2-form
$T^\a$ and the curvature 2-form $R^{\a\b}=-R^{\b\a}$. The first order
gravitational theory in this framework is called the Poincar\'e gauge
theory of gravity (PG).

The simplest model within PG is the Einstein-Cartan theory of gravity
(EC), see \cite{Trautman}, with the twisted gauge Lagrangian ($\kappa$
= gravitational and $\Lambda_0$ = cosmological
constant\footnote{Following essentially Schouten
    \cite{SchoutenRicci}, our conventions are \cite{PRs}: We have the
    coframe 1-form $\vt^\a=e_i{}^\a dx^i$ and the frame vectors
    $e_\b=e^j{}_\b\partial_j$, with $e_\b\rfloor\vt^\a=\d^\a_\b$. The
    connection 1-form is $\Gamma^{\a\b}=\Gamma_i{}^{\a\b}dx^i$. Greek
    indices are raised and lowered by means of the Minkowski metric
    $g_{\a\b}={\rm diag}(-1,1,1,1)$. The volume 4-form is denoted by
    $\eta$, and $\eta_\a={}^\star \vt_\a,\; \eta_{\a\b}={}^\star
    \vt_{\a\b},\; \eta_{\a\b\g}={}^\star \vt_{\a\b\g},\;
    \eta_{\a\b\g\d}={}^\star \vt_{\a\b\g\d}$, where $^\star$ is the
    Hodge star operator and $\vt^{\a\b}:=\vt^\a\wedge \vt^\b,\;{\rm
      etc}$. Furthermore, $ T^\a:=D\vt^\a=\frac 12\,T_{ij}{}^\a
    dx^i\wedge dx^j\,,\> T_{ij}{}^\a =
    2(\partial_{[i}\vt_{j]}{}^{\a}+\Gamma_{[ij]}{}^\a )\,,\>
    R^{\a\b}:=d\Gamma^{\a\b}+\Gamma^{\a\g}\wedge \Gamma^\b{}_\g=\frac
    12\,R_{ij}{}^{\a\b}dx^i\wedge dx^j\,,\>
    R_{ij}{}^{\a\b}=2(\partial_{[i}\Gamma_{j]}{}^{\a\b}
    +\Gamma_{[i}{}^{\a\g}\Gamma_{j]\,\cdot\,\g}^{\hspace{5pt}\b})\,,\>
    {\rm Ric}_\a:= e_\b\rfloor R_\a{}^\b={\rm
      Ric}_{\b\a}\,\vt^\b,\,{\rm with}\, {\rm
      Ric}_{\a\b}=R_{\g\a\b}{}^\g\,,\>R:={\rm
      Ric}_\a{}^\a=R_{\b\a}{}^{\a\b}\,.$ Diakonov et al.\
    \cite{Diakonov:2011fs} use the conventions of Landau-Lifschitz
    \cite{LL2nd}. Accordingly, there are the following correspondence
    rules: $R^\kappa{}_{\lambda\mu\nu}|_{\rm Diak\; et\;
      al.}={R_{\mu\nu\lambda}{}^{\kappa}|_{\rm here}}\,,\> {\rm
      Ric}_{\l\mu}|_{\rm Diak\; et\;
      al.}:=R_{\lambda}{}^{\kappa}{}_{\mu\kappa}|_{\rm Diak\; et\;
      al.}={ R_{\mu\kappa}{}^{\kappa}{}_{\lambda}|_{\rm here} =
      R_{\kappa\mu\lambda}{}^{\kappa}|_{\rm here}={\rm
        Ric}_{\mu\l}}|_{\rm here}$.}
\begin{equation}\label{VEC0}
\hspace{-35pt}  V_{{\rm EC}}:=
  \frac{1}{2\kappa}\left(\eta_{\a\b}\wedge
    R^{\a\b}-2\Lambda_0\eta\right) \quad{\rm and\>\,with}\quad
  L_{{\rm tot}}=V_{{\rm EC}}+ L(\psi,D\psi)\,,
\end{equation}
where $L$ is the matter Lagrangian depending on the minimally coupled
fermionic/bosonic matter fields $\psi(x)$. This is a viable
gravitational theory that deviates {}from general relativity at
extremely high matter densities $\rho\gtrsim \rho_{\rm crit}$, with
$\rho_{{\rm crit}}\approx m/\left(\lambda_{{\rm Compton}}\ell_{{\rm
      Planck}}^2 \right)$ and $m$ is the mass of the field, see also
\cite{Wei-Tou2010}. At the same time it is clear that GR can
alternatively be reformulated as a teleparallelism theory with torsion
square pieces in the Lagrangian. If we call the Newton-Einstein type
of gravity ``weak'' gravity, then its general quadratic gauge
Lagrangian reads ($a_0$ and $a_1,a_2,a_3$ are constants):
\begin{eqnarray}\label{V+weak}
\hspace{-35pt}  \label{weak}{V}^+_{\!{\rm weak}}  =
  \frac{1}{2\kappa}\left(-a_0R^{\alpha\beta}\wedge\eta_{\alpha\beta}
  -2\Lambda_{0}\eta  +T^\alpha\wedge\textstyle\sum
  \limits_{I=1}^{3}a_{I}{}^{\star(I)}
  T_\alpha\right)\,.
\end{eqnarray}
Here $^{(I)} T_\alpha$ denotes the irreducible pieces of the torsion,
with $^{(2)}T_\a:=\vt_\a\wedge(e_\b\rfloor T^\b)/3$ ({\tt vector}, 4
independent components), $^{(3)}T_\a:=e_\a\rfloor(T^\b\wedge \vt_\b)
/3$ ({\tt axitor}, 4), and $^{(1)}T_\a:=T_\a -{}^{(2)}T_\a -{}^{(3)}
T_\a$ ({\tt tentor}, 16). For the special cases $R^{\a\b}=0$, enforced
by a corresponding Lagrange multiplier term in (\ref{V+weak}), we
recover the teleparallel equivalent of GR, provided local Lorentz
invariance of the gravitational Lagrangian is implemented, see
\cite{Erice79,Mielke:1992te,Blagojevic:2000pi,Pereira:2001xf,Itin:2001bp,Obukhov:2002tm,Itin:2007sr},
and alternatively, for $T^\a=0$, we find GR directly. Thus, GR is
hidden in (\ref{V+weak}) in two totally different ways, a fact often
overlooked.

To link up with the experience of GR, we recall that the
Riemann-Cartan curvature 2-form ${R}^{\a\b}$ can be decomposed into
the (torsionfree) Riemann curvature $\widetilde{R}^{\a\b}$ and in
torsion dependent terms. For the curvature scalar this formula reads
(see \cite{PRs,Mielke:1992te})\footnote{The second minus sign on the
  right-hand-side of this equation is corrected. In \cite{PRs},
  Eq.(5.9.18) was a sign error.}
\begin{eqnarray} \label{App1}\hspace{-55pt}
 \hspace{-18pt} -{R}^{\alpha\beta}\wedge\eta_{\alpha\beta}=
  -\widetilde{R}^{\alpha\beta}\wedge\eta_{\alpha\beta}- T^{\alpha}\wedge
  {}^{\star}(-{}^{(1)}T_{\alpha} + 2{}^{(2)}T_{\alpha}
  +\frac 12{}^{(3)}T_{\alpha})+ 2d(\vartheta^{\alpha}\wedge
  {}^{\star}T_{\alpha})\,.
\end{eqnarray}
If we substituted (\ref{App1}) into (\ref{V+weak}), then apart from a
boundary term, see below, the Riemann curvature would emerge and the
$a_I$ in the quadratic torsion would get redefined. However, we don't
apply this procedure to (\ref{V+weak}), since we don't want to leave
the formalism of first order field theory.

We marked the Lagrangian (\ref{V+weak}) with a plus sign $+$ for
being, as a twisted 4-form, parity even. However, already in 1980,
Hojman, Mukku, and Sayed (HMS) \cite{Hojman:1980kv} and Nelson
\cite{Nelson:1980ph} added the parity odd\footnote{Parity even and
  {\em odd} torsion square terms were introduced by Purcell
  \cite{Purcell:1978zz} even 2 years ealier.} pseudoscalar curvature
piece $R_{\a\b}\wedge\vt^{\a\b}$ to the EC Lagrangian, see also
\cite{Nieh:1981ww,Bianchi,McCrea:1989sj,Baekler:1992,Obukhov:1995eq}. More
recently, in the context of the Ashtekar formalism \cite{Ashtekar},
see Kiefer \cite{Kiefer}, and in loop quantum gravity, see Rovelli
\cite{Rovelli,Rovelli:2011}, this has become popular, see
\cite{Holst:1995pc,Freidel:2005sn,Mercuri:2006um,Freidel:2006hv,Mercuri:2007ki,Bojowald:2007nu,Mukhopadhyaya:1998vp,Mukhopadhyaya:2002nu,Mielke:2009,Cantcheff,Date:2008rb,Mercuri:2009tz,Ertem,Wieland:2010ec,Daum:2010qt,Hanisch:2009xa,Pfaffle:2011,Pfaeffle:2011zu,Tilquin:2011bu,Kaul:2011va,Adak:2011md},
for related cosmological models see also
\cite{Puetzfeld:2004yg,Poplawski:2010jv,Poplawski:2010kb,Poplawski:2011xf,Randono:2010ym,Bjorken:2010qx}. Including
additionally odd torsion square pieces, we have
($b_0,\sigma_1,\sigma_2$ are constants, $^{(3)}\!R_{\alpha\beta}$ is
the irreducible pseudoscalar curvature 2-form)
\begin{eqnarray}\label{V-weak}
\hspace{-35pt}{V}^-_{\!{\rm weak}}  = -\frac{b_0}{2\kappa}\,^{(3)}\!R_{\alpha\beta}
  \wedge\vt^{\a\b}+ \displaystyle\frac{1}{{\kappa}}\left( {\sigma}_{1}{}
    ^{(1)}T^{\alpha}\wedge{} ^{(1)}T_{\alpha} + {\sigma}_{2}{}
    ^{(2)}T^{\alpha}\wedge{} ^{(3)}T_{\alpha}\right)\,.
\end{eqnarray}
The inverse of $b_0$ is sometimes called the Barbero-Immirzi parameter
\cite{Barbero:1994ap,Immirzi:1996di,Bojowald:2007nu}. The total gauge
Lagrangian would then read ${V}^+_{\!{\rm weak}}+{V}^-_{\!{\rm
    weak}}$. However, we should be aware that for weak gravity there
exists a boundary term, the untwisted parity odd {\it Nieh-Yan}
4-form\footnote{In a metric-affine spacetime \cite{PRs} with the
  distortion 1-form $N_{\a}{}^{\b}:= \Gamma_a{}^\b-
  \widetilde{\Gamma}_a{}^\b$, we can bring the Nieh-Yan identity in a
  very compact form, see \cite{Obukhov:1997pz}:
\begin{equation}
  d\left( T^{\a}\wedge {\vt}_{\a}\right)
  =\,  ^{(3)\!}R_{\a}{}^{\b}\wedge {\vt}^{\a}\wedge {\vt}_{\b} 
  - T^{\a}\wedge N_{\a}{}^{\b}\wedge {\vt}_{\b} \,.
\end{equation}} \cite{Nieh:1981ww},
\begin{eqnarray}\nonumber \hspace{-35pt}B^-_{TT}
=dC^-_{TT}& = &\displaystyle{1\over
    {2}}\,\left(T^{\alpha}\wedge T_{\a}+
    R_{\a\b}\wedge\vartheta ^{\a\b}\right)
  =\displaystyle {1\over {2}}(T^\a\wedge T_\a -{}^\star\! X) \\
\hspace{-35pt}  &=&\displaystyle {1\over {2}}\left(^{(1)}T^\a\wedge{} ^{(1)}T_\a
      +2\,^{(2)}T^\a\wedge {} ^{(3)}T_\a+ {}^{(3)}\!R_{\a\b}\wedge
      \vt^{\a\b}\right)\,,
\end{eqnarray}
with $C^-_{TT} :=\displaystyle\frac{1}{2}\vt^\a\wedge T_\a$ and $X$ as
the curvature pseudoscalar, $X=\eta_{\a\b\g\d}R^{[\a\b\g\d]}/4!$. We add
this form with a suitable constant $f_1$ to our weak gravity
Lagrangian:
\begin{equation}\label{LagrangianPM}
\hspace{-35pt}  V_{\!{\rm weak}}=V_{\!{\rm weak}}(a_0;b_0;a_1,a_2,a_3;\sigma_1,
  \sigma_2;f_1):= {V_{\!{\rm weak}}^{+}}+{V_{\!{\rm weak}}^{-}}
  + \frac{f_{1}}{\kappa}{B^-_{TT}}\,.
\end{equation}
It depends on the gravitational constant $\kappa$ and the cosmological
constant $\lambda_0$ and, furthermore, on the 8 constants specified in
(\ref{LagrangianPM}). By a suitable choice of $f_1$, we can compensate
either the HMS-term \cite{Hojman:1980kv} (that is, $b_0=0$) or one
tensor square term of the torsion (that is, either $\sigma_1=0$ or
$\sigma_2=0$). However, since $^{(1)}T^\a$ depends on 16 independent
components, the pseudoscalar curvature only on 1 component, it seems
to simplify the Lagrangian to a greater extent, if we kick out the
term with $^{(1)}T^\a$. Thus, for the weak gravity Lagrangian we are
left with {\bf 6} unspecified constants
${(a_0,b_0;a_1,a_2,a_3;\sigma_2)}$\footnote{Diakonov et al.\
  \cite{Diakonov:2011fs} found an {\it equivalent\/} result, but they
  eliminated the curvature pseudoscalar. Thus, they are left with the
  6 unspecified constants ${(a_0;a_1,a_2,a_3;\sigma_1,\sigma_2)}$,
  that is, with the curvature scalar plus the 5 torsion-square
  pieces.}.

Looking back at Eq.~(\ref{App1}), it could appear that we forgot the
boundary term $B^+_{TT}=\frac 12 d(\vt^\a \wedge{}^{\star}T_\a)$ and
that we could add it as $\frac{f_0}{\kappa}B^+_{TT}$ to the Lagrangian
(\ref{LagrangianPM}), see the procedure of Mielke
\cite{Mielke:2002sq,Mielke:2006gi}. However, if we compare the
Nieh-Yan and the ``teleparallel'' formulas
\begin{equation}\label{dT_N}
  \hspace{-65pt}  d\left({\vt}^{\a}\wedge T_{\a}\right) = T^{\a}
  \wedge T_{\a} - {\vt}^{\a}\wedge
  DT_{\a}\quad{\rm and}\quad d ( \vt^\a\wedge {}^\star T_\a)= T^\a \wedge
  {}^\star T_\a -  \vt^\a\wedge D \,^\star T_\a,
\end{equation}
respectively, then we recognize that in the former equation $DT_\a$
can be eliminated via the first Bianchi identity
$DT_\a=R_{\b\a}\wedge\vt^\b$, whereas in the latter equation such
  a trick is impossible. Therefore, we would trade in for torsion
  square pieces derivatives of the torsion and would mess up the
  first order character of our Lagrangian.\footnote
{We can combine the
    two equations in (\ref{dT_N}). This yields yields
\begin{equation}
  d\left({\vt}^{\a}\wedge {\xi}^{\pm}_{\a}\right) = T^{\a}\wedge
  {\xi}^{\pm}_{\a} - {\vt}^{\a}\wedge D{\xi}^{\pm}_{\a}\qquad
  {\rm with}\qquad {\xi}_{\a}^{\pm} := T_{\a} \pm \, ^{\star}T_{\a}\,.
\end{equation} Mielke \cite{Mielke:2002sq,Mielke:2006gi} built a 
similar linear combination but with the
imaginary unit in front of $^{\star}T_{\a}$, but neither his nor our
version seems to lead to firm conclusions up to now.}

If one starts with the EC-Lagrangian and adds {\it only} an HMS-term,
as numerous people do, then, because of the Nieh-Yan form, two torsion
square terms of odd parity are induced. Hence, the Lagrangian can be
reformulated as the EC-Lagrangian with specific additional torsion
square pieces:
\begin{equation}\label{newx}
 \hspace{-30pt} V_{\rm EC}+\frac{b_0}{2\kappa}\,^{(3)}\!R_{\alpha\beta}
  \wedge\vt^{\a\b}= V_{\rm EC}  
  -\frac{b_0}{2\kappa}\left(^{(1)}T^\a\wedge{} ^{(1)}T_\a
    +2\,^{(2)}T^\a\wedge {} ^{(3)}T_\a \right)+ d(\dots)\,.
\end{equation} 
Then the question can hardly be circumvented, why one should choose
only these specific torsion square pieces with very specific constants
and why the other torsion square pieces should be forbidden, that is,
the torsion square pieces come into focus. Moreover, it is known that
GR can be reformulated as a teleparallelism theory with torsion square
pieces in the Lagrangian
\cite{Erice79,Mielke:1992te,Blagojevic:2000pi,Pereira:2001xf,Itin:2001bp,Obukhov:2002tm,Itin:2007sr}. In
other words, the addition of the HMS-term opens the door wide for
torsion square Lagrangians.

Classically, it is consistent to consider only the two additional
specific terms in (\ref{newx}). However, it is not particularly
plausible. If torsion is introduced as a new concept, why should one
then introduce it in the highly constrained form of (\ref{newx})? In
loop quantum gravity \cite{Rovelli:2011}, which is thought of as a
fundamental theory of gravity, the truncated Lagrangian (\ref{newx})
is taken as a classical starting point, see \cite{Rovelli:2011}, Eq.\
(34), with the argument that also in QCD a similar parity odd piece is
used. However, then in the Lagrangian the internal color group $SU(3)$
with its potential $A$ is put in analogy to the local Poincar\'e group
${\rm R}^4\!\semidirect\! SO(1,3)$ with its translation potential
$\vt^\a$ and Lorentz potential $\Gamma^{\a\b}$. Apart from the fact
that QCD is quadratic in the field strength and (\ref{newx}) is only
linear in the curvature, this argument is less than convincing to us.

If gravity is seen in a quantum field theoretical context, see
Diakonov et al.\ \cite{Diakonov:2011fs}, for instance, then, as Date
et al.\ \cite{Date:2008rb} have pointed out, the Lagrangian
(\ref{newx}) is insufficient anyway: ``In a complete theory of
gravity, besides the Nieh-Yan topological term, we need to include two
other topological terms, the Pontryagin density and the Euler
density. This introduces two additional topological parameters
associated with such topological terms, besides the parameter ${\eta}$
[our $b_0$] we have discussed here. Any quantum theory of gravity
should have all these three CP-violating topological
couplings.''%\footnote{The topological Nieh-Yan term $B^{-}_{TT}$ and
 % the Pontryagin term $B^{-}_{RR}$ are both P odd and T odd; however,
 % the Euler term $B^{+}_{RR^{{\color{red}(*)}}}$ is P {\em even and T
%    even,} see also \cite{Kaul:2011va}. In \cite{Date:2008rb} it was
 % mentioned incorrectly that also the Euler term is P odd and T odd.}
 Actually, the Euler 4-form is CP-even, see equation (\ref{B_plus})
 or \cite{Kaul:2011va}.

From a totally different point of view, from observational cosmology
and from quantum chromodynamics, there are indications that we may
live in a parity violating Universe, see the review by Urban \&
Zhitnitsky \cite{Urban:2010wa}. All the more investigations in a
parity odd PG model seem desirable.

%%%%%%%%%%%%%%%%%%%%%%%%%%%%%%%%%%%%%%%%%%%%%%%%%%%%%%%%%%%%%%%%%%%%%%%%
\section{Quadratic Poincar\'e gauge theory and strong gravity}
%%%%%%%%%%%%%%%%%%%%%%%%%%%%%%%%%%%%%%%%%%%%%%%%%%%%%%%%%%%%%%%%%%%%%%%%

If one wants the Lorentz connection as a propagating field, then one
has to allow for ``strong'' gravity of the Yang-Mills type by adding
quadratic curvature 4-forms to the weak Lagrangian. The curvature
$R^{\a\b}$ of a Riemann-Cartan space has six irreducible pieces:
$R^{\a\b}=\textstyle\sum_{I=1}^6{} ^{(I)}\!R^{\a\b}$. We write
symbolically, using the self-explanatory computer names for the
irreducible terms: {\tt curv} (36 indep.\ comp.) = {\tt weyl} (10) +
{\tt paircom} (9) + {\tt pscalar} (1) + {\tt ricsymf} (9) + {\tt
  ricanti} (6) + {\tt scalar} (1), see \cite{PRs} for the exact
definitions. In a Riemann space only $^{(1)}\!R^{\a\b}$,
$^{(4)}\!R^{\a\b}$, and $^{(6)}\!R^{\a\b}$ are left over. Hence the
most general parity even quadratic Lagrangian, with a new
dimensionless coupling constant $\varrho$, reads
\begin{eqnarray} {V}^+_{\!{\rm strong}} =
  -\frac{1}{2\varrho}R^{\alpha\beta}
  \wedge{}\textstyle\sum\limits_{I=1}^{6}w_{I}\,^{\star(I)}\!
  R_{\alpha\beta}\,.
\end{eqnarray}
Alerted by the corresponding case in weak gravity, we now search for
parity odd terms. They were found to be (see
\cite{YuriEtAl,Baekler:2010fr}) as
\begin{eqnarray}
 {V}^-_{\!{\rm strong}} &=& \nonumber
-\frac{1}{2{\varrho}}\left( {\mu}_{1}{} ^{(1)}\!R^{\alpha\beta}\wedge{}
    ^{(1)}\!R_{\alpha\beta} + {\mu}_{2}{} ^{(2)}\!R^{\alpha\beta}\wedge{}
    ^{(4)}\!R_{\alpha\beta}\right. \\ & &\hspace{18pt}
  + \left.
    {\mu}_{3}{}^{(3)}\!R^{\alpha\beta}\wedge{} ^{(6)}\!R_{\alpha\beta} +
    {\mu}_{4}{} ^{(5)}\!R^{\alpha\beta}\wedge{} ^{(5)}\!R_{\alpha\beta}
  \right),
\end{eqnarray}
which are the only quadratic curvature square invariants of odd
character in a 4D Riemann-Cartan space. Note that in a Riemann space,
that is, when torsion vanishes, only the first piece built up {}from the
Weyl curvature $^{(1)}\!R^{\a\b}$ is left over.

Taking a lesson {}from the above, we can now search for boundary
terms. As in any Yang-Mills theory, we can find an untwisted {\it
  Pontryagin} 4-form $B^-_{RR}$. But in gravity the (anholonomic)
Lorentz indices of the curvature can be contracted with the help of
the totally antisymmetric Levi-Civita tensor
$\eta^{\a\b\g\d}$. Accordingly, we introduce the so-called Lie-dual of
the curvature with the Lie star operator $^\rstar$ as
\begin{equation}
R^{\rstar\, \a\b} := \frac{1}{2}R_{\mu\nu}{\eta}^{\mu\nu\a\b}\,.
\end{equation}
Like ${\eta}^{\mu\nu\a\b}$, the Lie star $^ \rstar$ is twisted. It
gives rise to the twisted {\it Euler} 4-form $B^{+}_{RR^{\rstar
  }}$. Following \cite{PRs}, we have then the following two boundary
terms:
\begin{equation}\label{BRRBRR*}
\hspace{-40pt}  B^-_{RR}=dC^-_{RR}={1\over 2}R_{\alpha\beta}\wedge
  R^{\a\beta}\,,\quad B^{+}_{RR^{\rstar }}=dC^+_{RR^{\rstar}}
  ={1\over 2}R_{\a\b} \wedge
  R^{\rstar\a\b}\,.
\end{equation}
Thus, the strong part of the gauge Lagrangian turns out to be
\begin{equation}\label{LagrangianPM1}
  \hspace{-40pt}  V_{\!{\rm strong}}:= {V_{\!{\rm strong}}^{+}}
  +{V_{\!{\rm strong}}^{-}}
  +\frac{f_{2}}{\varrho}{B^-_{RR}}+
  \frac{f_{3}}{\varrho} {B^+_{RR^{\rstar}}}\,.
\end{equation}
The total quadratic gauge Lagrangian including boundary terms is then
\begin{equation}
\hspace{-40pt}  V_{\rm gauge} =  {V_{\!{\rm weak}}^{+}}+{V_{\!{\rm weak}}^{-}}
  +{V_{\!{\rm strong}}^{+}}
  +{V_{\!{\rm strong}}^{-}} + \frac{f_{1}}{\kappa}{B^{-}_{TT}}
  + \frac{f_{2}}{\varrho}{B^-_{RR}}+
    \frac{f_{3}}{\varrho} {B^+_{RR^{\rstar}}}.
\end{equation}
For {\it vanishing torsion,} $V_{{\rm weak}}$ reduces to $V_{\rm GR}$
with cosmological constant and $V_{{\rm strong}}$ has only
$(w_1,w_4,w_6;\mu_1;f_2,f_3)\ne 0$. By a suitable choice of $f_2,f_3$,
only the two terms with $w_4,w_6$ survive, that is, those with the
tracefree Ricci tensor and the curvature scalar. 

%%%%%%%%%%%%%%%%%%%%%%%%%%%%%%%%%%%%%%%%%%%%%%%%%%%%%%%%%%%%%%%%%%%%%%%%
\section{The role of the Lie-dual of the curvature}
%%%%%%%%%%%%%%%%%%%%%%%%%%%%%%%%%%%%%%%%%%%%%%%%%%%%%%%%%%%%%%%%%%%%%%%%

Before we continue the investigation of (\ref{LagrangianPM1}), we will
derive some rules for manipulating curvature square terms containing a
Lie star. Using heavily the computer-algebra system {\tt Reduce} with
the {\tt Excalc} package, compare
\cite{Hearn,Schruefer,EXCALC,Socorro:1998hr}, we were able to convert
completely the Lie star $^\rstar$ into the Hodge star $^\star$ according
to the following rules: The expression $^{(I)}\!R^{\mu\nu}\wedge
^{(J)}\!R^{\rstar}_{\mu\nu}$ is diagonal, that is, $\propto
{\delta}^{IJ}$; only the diagonal pieces do not vanish, namely
\begin{equation}
  ^{(I)}\!R^{\mu\nu}\!\wedge{} ^{(I)}\!R^{\rstar}_{\mu\nu}=
  \pm\,^{(I)}\!R^{\mu\nu}\!\wedge{} ^{\star(I)}\!R_{\mu\nu}\,,
\end{equation}
with $+$ for $I=1,3,5,6$ and with $-$ for $I=2,4$. Note that on the
left-hand-side of this equation we have the Lie star $^\rstar$, on the
right-hand-side, however, the Hodge star $^\star$.  This implies the
relation, derived here for the first time explicitly\footnote{By using
  some simple algebra, Eq.(\ref{new}) can alternatively be derived in
  a straightforward way from Eqs.(10.17) to (10.22) of Obukhov
  \cite{Obukhov:2006gea}.},
\begin{eqnarray}\label{RR_04*}
%\begin{tabular}{|c|}\hline\cr
\hspace{-35pt}  R^{\a\b}\wedge R^{\rstar}_{\a\b} &=& {} ^{(1)}\!R^{\a\b}\wedge{}
  ^{\star(1)}\!R_{\a\b} -{} ^{(2)}\!R^{\a\b}\wedge{} ^{\star
    (2)}\!R_{a\b} +{} ^{(3)}\!R^{\a\b}\wedge{} ^{\star
    (3)}\!R_{a\b}\cr\hspace{-35pt} &&\hspace{-13pt}-{}^{(4)}\!R^{\a\b}\wedge
  {}^{\star(4)}\!R_{\a\b} +{} ^{(5)}\!R^{\a\b}\wedge{} ^{\star
    (5)}\!R_{a\b} +{} ^{(6)}\!R^{\a\b}\wedge{} ^{\star(6)}\!R_{a\b}\,.\label{new}
%\cr\hline\end{tabular}
\end{eqnarray}
In particular, this shows that the Lie star is superfluous in forming
a quadratic Lagrangian, the Hodge star is sufficient.

Comparison with (\ref{BRRBRR*}) allows us to rewrite the {\it Euler}
4-form with the help of the Hodge star as
\begin{eqnarray}\label{B_plus}
\hspace{-35pt}B^+_{RR^{\rstar}}=\frac{1}{2}\left(R^{\a\b}\wedge{}^\star R_{\a\b}
-2{}^{(4)}\!R^{\a\b}\wedge{} ^{\star (4)}\!R_{\a\b}-2
^{(2)}\!R^{\a\b}\wedge{} ^{\star (2)}\!R_{\a\b}\right)\,.
\end{eqnarray}
The {\it Pontryagin} 4-form, also defined in (\ref{BRRBRR*}), after some
algebra, can be expressed in terms of the irreducible pieces of
the curvature as (also this relation is new)
\begin{eqnarray}\label{Pontryagin}
\hspace{-35pt} B^-_{RR}&=&\frac{1}{2}\left(  \,^{(1)}\!R_{\a\b}\wedge{}
  ^{(1)}\!R^{\a\b}+\;\,^{(5)}\!R_{\a\b}\wedge{}^{(5)}\!R^{\a\b}\right.
  \cr
  & &\left. +2\,^{(3)}\!R_{\a\b} \wedge{}^{(6)}\!R^{\a\b}
  +2\,^{(2)}\!R_{\a\b} \wedge{}^{(4)}\!R^{\a\b}\right)\,.
\end{eqnarray}

\section{Comparison with Diakonov et al. \cite{Diakonov:2011fs}}

Recently, in the framework of perturbative quantum field theory, new
results were reported \cite{Diakonov:2011fs} on including torsion in a
gravitational gauge theory for describing fermionic matter, for a
review of some earlier results, see \cite{Shapiro:2001rz}. In this
context, Diakonov et al.\ investigated gravitational gauge Lagrangians
containing quadratic terms in the gauge fields of even and of odd
parity. That is, contributions of the weak and the strong gravity
sector (Lorentz gauge bosons) were considered. Generally, those
4-fermion interaction terms will give additional contributions to the
energy-momentum current of matter. This aspects might be relevant in
the context of quantum cosmological models, see
\cite{Poplawski:2010jv,Randono:2010ym,Bjorken:2010qx}.  Other groups
address only weak gravity, even though they include Euler and
Pontryagin terms, which refer to strong gravity, see Benedetti et al.\
\cite{Benedetti:2011nd}.

In the following, we would like to compare the approach given in
\cite{Diakonov:2011fs} with the results we already gave in
\cite{Baekler:2010fr}.

\subsection{Torsion square invariants}

Let us compare our torsion-square invariants in (\ref{LagrangianPM}),
see also \cite{Baekler:2010fr}, with those in \cite{Diakonov:2011fs},
Eq.(53). If we add a plus sign for parity even and a minus sign for
parity odd terms, the invariants of Diakonov et al., multiplied by the
volume form $\eta$, read:
\begin{eqnarray}\label{2}
  K_{1}^+ & = & 2\, ^{(1)}T^{\a}\wedge\, ^{\star (1)}T_{\a} 
  - 2\, ^{(2)}T^{\a}\wedge\, ^{\star (2)}T_{\a}
  + 2\, ^{(3)}T^{\a}\wedge\, ^{\star (3)}T_{\a}\,,\\
  K_{2}^+ & = & 3\, ^{(2)}T^{\a}\wedge\, ^{\star (2)}T_{\a}\,,\\
  K_{3}^+ & = & \, ^{(1)}T^{\a}\wedge\, ^{\star (1)}T_{\a} 
  + \, ^{(2)}T^{\a}\wedge\, ^{\star (2)}T_{\a}
  - 2\, ^{(3)}T^{\a}\wedge\, ^{\star (3)}T_{\a}\,,\\
  K_{4}^- & = & -2\,^{(1)}T^{\a}\wedge\, ^{(1)}T_{\a} 
  + 4\, ^{(2)}T^{\a}\wedge\, ^{(3)}T_{\a}\,,\\
  K_{5}^- & = & \,^{(1)}T^{\a}\wedge\, ^{(1)}T_{\a} 
  + 4\, ^{(2)}T^{\a}\wedge\, ^{(3)}T_{\a}\,,
\end{eqnarray}
and the inverse relations are
\begin{eqnarray}
  ^{(1)}T^{\a}\wedge\, ^{\star (1)}T_{\a} & = &
  \frac{1}{9}(3K_{1}^++K_{2}^+  +3K_{3}^+)\,,\\
  ^{(2)}T^{\a}\wedge\, ^{\star (2)}T_{\a} & = & \frac{1}{3}K_{2}^+\,, \\
  ^{(3)}T^{\a}\wedge\, ^{\star (3)}T_{\a} & = & \frac{1}{12}(3K_{1}^+
  +4K_{2}^+-6K_{3}^+)\,,\\
  ^{(1)}T^{\a}\wedge\, ^{(1)}T_{\a} & = & -\frac{1}{3}(K_{4}^--K_{5}^-)\,,\\
  ^{(2)}T^{\a}\wedge\, ^{(3)}T_{\a} & = & \frac{1}{12}(K_{4}^-+2K_{5}^-)\,.
\end{eqnarray}
These 5 invariants agree with those given in \cite{Baekler:2010fr},
Eqs.(30) and (55).

%--------------------------------------------------------------------
\subsection{Curvature square invariants}

Diakonov et al.\ \cite{Diakonov:2011fs}, Eq.\ (59), find 6 even and 4 odd
independent quadratic invariants $G_I$. As we did with the torsion
invariants, we translate their component representations into the
language of exterior differential forms used by us. We find, after
some messy computer checked algebra, the following curvature
invariants (multiplied by $\eta$):
\begin{eqnarray}\label{G_K}
  \hspace{-20pt}G^+_1 & = & R^2{\eta} = 12\,^{(6)}\!R_{\a\b}\wedge
  {}^{\star(6)}\!R^{\a\b}  \,,\\
  \hspace{-20pt}G^+_2 & = & R_{\mu\nu\r\l}R^{\mu\nu\r\l}{\eta} =
  2\,R_{\a\b}\wedge{}^{\star}R^{\a\b} = 2\sum
  \limits_{I=1}^{6}{}^{(I)}\!R^{\alpha\beta}
  \wedge{}^{\star(I)}\!
  R_{\alpha\beta}\,,\\
  \hspace{-20pt}G^+_3 & = & R_{\mu\nu\r\l}
  R^{\lambda\rho\nu\mu}{\eta}\cr
  &  =& 2\left(^{(1)}\!R^{\a\b}\wedge{}^{\star
      (1)}\!R_{\a\b}-{}^{(2)}\!R^{\a\b}
    \wedge{} ^{\star (2)}\!R_{\a\b}
    + {}^{(3)}\!R^{\a\b}\wedge{} ^{\star (3)}\!R_{\a\b}\right. \cr
  & & \left. +{}^{(4)}\!R^{\a\b}\wedge{} ^{\star (4)}\!R_{\a\b}
    -{}^{(5)}\!R^{\a\b}\wedge{} ^{\star (5)}\!R_{\a\b}
    + {}^{(6)}\!R^{\a\b}\wedge{} ^{\star (6)}\!R_{\a\b}\right)
  \,,\\
  \hspace{-20pt}G^+_4 & = & \left( R^2
    - 4{\rm Ric}_{\mu\lambda}{\rm Ric}^{\lambda\mu}
    +R_{\mu\nu\r\l}R^{\lambda\rho\nu\mu}\right){\eta}\cr
  &= & 2\left(^{(1)}\!R^{\a\b}\wedge{} ^{\star (1)}\!R_{\a\b}
    -{}^{(2)}\!R^{\a\b}\wedge{} ^{\star (2)}\!R_{\a\b}
    +{} ^{(3)}\!R^{\a\b}\wedge{} ^{\star (3)}\!R_{\a\b}\right. \cr
  & & \left.\, -{} ^{(4)}\!R^{\a\b}\wedge{} ^{\star (4)}\!R_{\a\b}
    +{}^{(5)}\!R^{\a\b}\wedge{} ^{\star (5)}\!R_{\a\b}
    +{} ^{(6)}\!R^{\a\b}\wedge{} ^{\star (6)}\!R_{\a\b}\right)\,,\\
  \hspace{-20pt}G^+_5 & = & \left( R^2 - 4{\rm Ric}_{\mu\lambda}
    {\rm Ric}^{\mu\lambda}
    +R_{\mu\nu\r\l}R^{\mu\nu\r\l}\right){\eta}\cr
  &= &  2\left(^{(1)}\!R^{\a\b}\wedge{} ^{\star (1)}\!R_{\a\b}+{}
    ^{(2)}\!R^{\a\b}\wedge{} ^{\star (2)}\!R_{\a\b}
    +{} ^{(3)}\!R^{\a\b}\wedge{} ^{\star (3)}\!R_{\a\b}\right. \cr
  & & \left.  -{} ^{(4)}\!R^{\a\b}\wedge{} ^{\star(4)}\!R_{\a\b}
    -{} ^{(5)}\!R^{\a\b}\wedge{}^{\star (5)}\!R_{\a\b}
    +{} ^{(6)}\!R^{\a\b}\wedge{} ^{\star (6)}\!R_{\a\b}\right)
  \,,\\
  \hspace{-20pt}G^+_6 & = & \left(
    {\eta}^{\lambda\rho\mu\nu}R_{\mu\nu\r\l}\right)^{2}
  {\eta} = -48\,^{(3)}\!R_{\a\b}\wedge{}^{\star(3)}\!R^{\a\b}\,,\\
  \hspace{-20pt}G^-_7& = &
  R{\eta}^{\lambda\rho\mu\nu}R_{\mu\nu\r\l}{\eta}
  = -24\,^{(3)}\!R_{\a\b}
  \wedge{}^{(6)}\!R^{\a\b}\,,\\
  \hspace{-20pt}G^-_8 & = & \eta^{\mu\nu\a\b}R_{\mu\nu\r\l}
  R_{\a\b}{}^{\r\l}\eta \cr & = &  -4\,^{(1)}\!R_{\a\b}\wedge{}
  ^{(1)}\!R^{\a\b}-4\,^{(5)}\!R_{\a\b}\wedge{}^{(5)}\!R^{\a\b}
  \cr
  & & -8\,^{(3)}\!R_{\a\b} \wedge{}^{(6)}\!R^{\a\b}
  -8\,^{(2)}\!R_{\a\b} \wedge{}^{(4)}\!R^{\a\b}\,,\\
  \hspace{-20pt}G^-_9 & = & \eta^{\lambda\rho\g\d}R_{\mu\nu\r\l}
  R^{\mu\nu}{}_{\d\g}\eta\cr & = &  -4\,^{(1)}\!R_{\a\b}\wedge{}
  ^{(1)}\!R^{\a\b}-4\,^{(5)}\!R_{\a\b}\wedge{}^{(5)}\!R^{\a\b}
  \cr
  & &-8\,^{(3)}\!R_{\a\b} \wedge{}^{(6)}\!R^{\a\b}
  +8\,^{(2)}\!R_{\a\b} \wedge{}^{(4)}\!R^{\a\b}\,,\\
  \hspace{-20pt}G^-_{10} & = & \eta^{\lambda\rho\a\b}\label{last}
  R_{\mu\nu\r\l}
  R_{\a\b}{}^{\nu\mu}\eta \cr & = & -4\,^{(1)}\!R_{\a\b}\wedge{}
  ^{(1)}\!R^{\a\b}+4\,^{(5)}\!R_{\a\b}\wedge{}^{(5)}\!R^{\a\b}
  -8\,^{(3)}\!R_{\a\b} \wedge{}^{(6)}\!R^{\a\b}\,.\end{eqnarray}
The inverse relations are convenient for a detailed comparison. They
turn out to be
\begin{eqnarray}
  \hspace{-20pt}  ^{(1)}\!R^{\a\b}\wedge{} ^{\star (1)}\!R_{\a\b} 
  & = & -\frac{1}{12}G_{1}^{+}
  +\frac{1}{8}\left(G_{2}^{+}+G_{3}^{+}+G_{4}^{+}+G_{5}^{+}\right)
  + \frac{1}{48}G_{6}^{+}\,,\\ & & \cr
 \hspace{-20pt}  ^{(2)}\!R^{\a\b}\wedge{} ^{\star (2)}\!R_{\a\b} 
  & = & \frac{1}{8}\left(G_{2}^{+}-G_{3}^{+}-G_{4}^{+}
    +G_{5}^{+}\right)\,,\\ & & \cr
  \hspace{-20pt} ^{(3)}\!R^{\a\b}\wedge{} ^{\star (3)}\!R_{\a\b} & = & 
  -\frac{1}{48}G_{6}^{+}\,,\\ & & \cr
  \hspace{-20pt} ^{(4)}\!R^{\a\b}\wedge{} ^{\star (4)}\!R_{\a\b} & = & 
  \frac{1}{8}\left(G_{2}^{+}+G_{3}^{+}
    -G_{4}^{+}-G_{5}^{+}\right)\,,\\ & & \cr
  \hspace{-20pt} ^{(5)}\!R^{\a\b}\wedge{} ^{\star (5)}\!R_{\a\b} & = & 
  \frac{1}{8}\left(G_{2}^{+}-G_{3}^{+}
    +G_{4}^{+}-G_{5}^{+}\right)\,,\\ & & \cr
  \hspace{-20pt} ^{(6)}\!R^{\a\b}\wedge{} ^{\star (6)}\!R_{\a\b} & = & 
  \frac{1}{12}G_{1}^{+}\,,
\end{eqnarray}
and
\begin{eqnarray}
 \hspace{-20pt}  ^{(1)}\!R^{\a\b}\wedge{} ^{(1)}\!R_{\a\b} & = & 
  -\frac{1}{16}\left(G_{8}^{-}+G_{9}^{-}+2G_{10}^{-}\right)
  +\frac{1}{12}G_{7}^{-}\,,\\ & & \cr
  \hspace{-20pt} ^{(2)}\!R^{\a\b}\wedge{} ^{(4)}\!R_{\a\b} & = & 
  -\frac{1}{16}\left(G_{8}^{-}-G_{9}^{-}\right)\,,\\ & & \cr
   \hspace{-20pt}^{(3)}\!R^{\a\b}\wedge{} ^{(6)}\!R_{\a\b} & = & 
  -\frac{1}{24}G_{7}^{-}\,,\\ & & \cr
  \hspace{-20pt} ^{(5)}\!R^{\a\b}\wedge{} ^{(5)}\!R_{\a\b} & = & 
  -\frac{1}{16}\left(G_{8}^{-}+G_{9}^{-}-2G_{10}^{-}\right)\,.
\end{eqnarray}

It is now straightforward to express the Euler 4-form (\ref{B_plus})
and the Pontryagin 4-form (\ref{Pontryagin}) in terms of the
$G_I$'s. We find
\begin{equation}\label{Euler_G}
B_{RR^{\rstar}}^{+} = \frac{1}{4}G_{4}^{+}\qquad{\rm and}\qquad 
B_{RR}^{-} = -\frac{1}{8}G_{8}^{-}\,,
\end{equation}
respectively. This is what Diakonov et al.\ stressed: that their
invariants $G_{4}^{+}$ and $G_{8}^{-}$ are boundary terms. These two
boundary terms can also be found in our earlier work, see
\cite{Baekler:2010fr}, Eqs.(33) and (50). 

Hence the results of Diakonov et al.\ \cite{Diakonov:2011fs} with
respect to the quadratic invariants of torsion and curvature coincide
with those of \cite{Baekler:2010fr}. This is also manifest in the
Riemannian subcase, that is, for {\it vanishing torsion}
$T^\a=0$. Then, 
\begin{equation}\label{V4}
^{(2)}\!R_{\a\b}={} ^{(3)}\!R_{\a\b}={}
^{(5)}\!R_{\a\b}=0\,,
\end{equation} or, in terms of the $G_I$'s,
\begin{equation}\label{Riemann}
   G_{2}^{+}=G_{3}^{+}\,, \quad G_{4}^{+}=G_{5}^{+}\,, \quad
   G_{6}^{+}=G_{7}^{+}=0\,,\quad G_{8}^{+}=G_{9}^{+}=G_{10}^{+}\,,
\end{equation}
which can be read off directly {}from the Eqs.(\ref{G_K}) to
(\ref{last}). Under the condition of vanishing torsion the boundary
terms read
\begin{eqnarray}
 \hspace{-50pt}   B^+_{RR^{\rstar}}|_{T^{\a}=0} &=&  
  \frac 12\left({} ^{(1)}\!R^{\a\b}\wedge{}
    ^{\star(1)}\!R_{\a\b}-{}^{(4)}\!R^{\a\b}\wedge
    {}^{\star(4)}\!R_{\a\b} 
    +{} ^{(6)}\!R^{\a\b}\wedge{} ^{\star(6)}\!R_{a\b}\right),\\
 \hspace{-50pt}   B_{RR}^{-}|_{T^{\a}=0} &=& \frac 12\, ^{(1)}R^{\a\b}\!\wedge 
  \,\! ^{(1)}\!R_{\a\b}\,.\label{XXX}
\end{eqnarray}

%%%%%%%%%%%%%%%%%%%%%%%%%%%%%%%%%%%%%%%%%%%%%%%%%%%%%%%%%%%%%%%%%%%%%%%%
\section{The number of independent terms in the most general
quadratic PG Lagrangian}
%%%%%%%%%%%%%%%%%%%%%%%%%%%%%%%%%%%%%%%%%%%%%%%%%%%%%%%%%%%%%%%%%%%%%%%%

For the strong gravitational Lagrangian $V_{\!{\rm strong}}$ in
(\ref{LagrangianPM1}), we can enter in a similar discussion as for
$V_{\!{\rm weak}}$ in (\ref{LagrangianPM}): Besides the strong
gravitational coupling constant $\varrho$, we have the 12 constants
${(w_1,w_2,...,w_6;\mu_1,\mu_2,\mu_3,\mu_4;f_2,f_3)}$. By a suitable
choice of $f_2$ and $f_3$, we can compensate the terms containing the
Weyl 2-form $^{(1)}\!R^{\a\b}$ with 10 independent components, as can
be seen {}from (\ref{B_plus}) and (\ref{Pontryagin}). Consequently, we
are left with the {\bf 8} constants
${(w_2,...,w_6;\mu_2,\mu_3,\mu_4)}$. Diakonov et al.\
\cite{Diakonov:2011fs} found the 10 invariants $G_I$, two of which,
namely $G^+_4$ and $G_8^-$ are boundary terms. Hence they also arrive
at 8 independent invariants. Accordingly, also for strong gravity our
results match those of Diakonov et al.

Our final gravitational Lagrangian is then\footnote{If we introduce
  the notations $R$ and $X$ for the curvature scalar and the curvature
  pseudoscalar, then we find $^{(6)}R_{\a\b} = -R\,\vt_{\a\b}/12$ and
  $^{(3)}R_{\a\b}= -X\eta_{\a\b}/12$, respectively; moreover, for the
  torsion we can define the 1-forms of $\cal A$ and $\cal V$ for the
  axial vector and the vector torsion $^{(3)}T^{\a} = {}^{\star}({\cal
    A}\wedge\, {\vartheta}^{\a})/3$ and $^{(2)}T^\a=-({\cal
    V}\wedge\vt^\a)/3$, respectively.}
\begin{eqnarray} \label{QMA}\nonumber \hspace{-50pt} V &=&
  \,\frac{1}{2\kappa}[\,\left(\,a_0R-2\Lambda_{0}+{b_0}X)\,\eta\right.\\
  \hspace{-50pt} && \hspace{13pt}\left. +\frac{a_{2}}{3} {\cal V}\wedge
    {}^{\star\!} {\cal V} -\frac{a_{3}}{3}{{\cal A} \wedge{}
      ^{\star\!\!\!}{\cal A}} -\frac{2{\sigma}_{2}}{3}{ {\cal
        V}\wedge{} ^{\star\!\!\!} {\cal A}}+ a_{1}{}^{(1)}T^\alpha
    \wedge
    {}^{\star(1)}T_\a \right]\nonumber \\
  \hspace{-50pt} & &\hspace{-11pt} -\frac{1}{2\varrho} \left[
    (\frac{w_6}{12} R^2- \frac{w_3}{12} {X^2} + \frac{\mu_3}{12}
      R X)\,\eta + w_{4}{}^{(4)}\!R^{\a\b}\wedge
    {}^{\star(4)}\!R_{\a\b}\right.\nonumber\cr \hspace{-50pt} && \\
  \hspace{-50pt} & &\left. +{}^{(2)}\!R^{\a\b}\wedge (
    w_2{}^{\star(2)}\!R_{\a\b}
    +\mu_2{}^{(4)}\!R_{\a\b})+{}^{(5)}\!R^{\a\b}
    \wedge(w_5{}^{\star(5)}\!R_{\a\b}
    +\mu_4{}^{(5)}\!R_{\a\b})\right].
\end{eqnarray}
The first two lines represent weak gravity, the last two lines strong
gravity. The parity odd pieces are those with the constants
$b_0,\sigma_2;\mu_2,\mu_3,\mu_4$. In a Riemann space (where $X=0$),
only two terms of the first line and likewise two terms in the third
line survive. All these 4 terms are parity even, that is, only torsion
brings in parity odd pieces into the gravitational Lagrangian.

Yo and Nester \cite{Yo:1999ex,Yo:2001sy,Yo:2006qs} found that only a
small subclass of the Lagrangians (\ref{QMA}) is consistent {}from a
Hamiltonian point of view. They, together with Shie, presented such a
Lagrangian \cite{Shie:2008ms} and found an accelerating cosmological
Friedman type model with propagating connection. Shortly afterwards,
Nester and his group, see Chen et al.\ \cite{exSNY}, generalized this
model and introduced a consistent Lagrangian containing the parity odd
pieces $\cal A$ and $X$. Since these terms occur quadratically, their
Lagrangian was still parity even:
\begin{eqnarray}\label{extendedSNY}
  \hspace{-50pt} V_{{\rm Chen\> et\> al.}}&=&\;
  \frac{1}{2\kappa}\left(a_0 R-2\Lambda_0\right)\eta
  +\frac{1}{6\kappa}\left(a_{2}  {\cal V}\wedge {}^\star  {\cal V}-
    a_{3}{ {\cal A}\wedge {}^{\star\!\!\!} {\cal A}}\right)
  \nonumber\\\hspace{-50pt} &&\hspace{-10pt}
  -\frac{1}{24\varrho}\left(w_6R^2-w_3{X^2}\right)\eta\,.
\end{eqnarray}
The next step was done by Baekler et al.\ \cite{Baekler:2010fr}, see
also \cite{BianchiBHN,Ho:2011xf}. They investigated a Lagrangian with
three additional pieces with odd parity ({\ref{extendedSNY}}), namely
those carrying the constants $b_0,\sigma_2,\mu_3$:
\begin{eqnarray}\label{L_BHN'}
 \hspace{-50pt} V_{\rm BHN} & = &\;
  \frac{1}{2{\kappa}}(a_{0}R -2{\Lambda}_{0}+b_{0}{X}){\eta}
  +\frac{1}{6{\kappa}}\left(a_{2} {\cal V}\wedge
    {}^{\star\!} {\cal V} -a_{3}{{\cal A}\wedge{} ^{\star\!\!\!}{\cal A}}
    -2{\sigma}_{2}{ {\cal V}\wedge{}
      ^{\star\!\!\!} {\cal A}}\right)\nonumber\\
\hspace{-50pt}  &&\hspace{-7pt}-\frac{1}{24{\varrho}}\left( w_{6}
R^2- w_{3} {X^2} + {\mu}_{3}R X\right)\eta\,.
\end{eqnarray}
Now one should analyze the particle content of the lagrangian
(\ref{QMA}), that is, to find out which modes are propagating
decently. This has been done in \cite{Baekler:2010fr} by the simple
method of the {\it diagonalization of the Lagrangian}. The results
turned out to be in agreement with those of the Hamiltonian approach.

It is manifest already by now, looking beyond the Einstein-Cartan
theory including parity odd Lagrangians is a field with bright
prospects.

\ack \vspace{-10pt}
\begin{footnotesize}
  We are very grateful to Dmitri Diakonov (St.~Petersburg) and his
  collaborators, in particular also to Alexander Tumanov
  (St.~Petersburg), for many helpful email discussions on the content
  of their paper. We thank Yuri Obukhov (London/Moscow) for numerous
  useful remarks and for pointing out the relevance of Ref.\
  \cite{Purcell:1978zz}. We are also grateful to Yakov Itin
  (Jerusalem) and James Nester (Chung-li) for helpful remarks and to
  Ghanashyam Date (Chennai) for drawing our attention to Ref.\
  \cite{Date:2008rb}. Eckehard Mielke (Mexico City) pointed out to us
  the importance of the boundary term in Eq.\ (\ref{App1}). Many
  thanks to him and also, last but not least, to J.~D.~Bjorken (SLAC)
  for alerting us to a theory of a possible parity odd
  Universe. P.~B. would like to acknowledge support of FB Medien at
  FH-D.
\end{footnotesize}

\end{document}